\documentclass[cits]{PoS}
\usepackage{graphicx}
\usepackage{amsfonts}
\usepackage{dsfont}
\usepackage[intlimits]{amsmath}
\usepackage{latexsym}
\usepackage{bm}

\title{Thermodynamical quantities for overlap fermions with chemical
potential}

\ShortTitle{Thermodynamical quantities for overlap fermions}

\author{Christof Gattringer$^a$ and  \speaker{Ludovit Liptak}\thanks{Our research is
supported in part by the Slovak Science and Technology Assistance Agency, Grant No.\ APVT--51--005704, and the Grant Agency for Science, Project VEGA No.\ 2/6068/2006 (L.L.).
Participation of the speaker at the conference was enabled by support of the ESF project CEPOS at FMFI of the Comenius Univ.\ in Bratislava.} $^b$\\
\llap{$^a$}  Institut f\"ur Physik, Universit\"at Graz, 
Universit\"atsplatz 5, 8010 Graz, Austria\\
\llap{$^b$} Institute of Physics, Slovak Academy of Sciences, 
D\'ubravsk\'a cesta 9, 845 11 Bratislava 45, Slovak Republic\\

E-mail: \email{christof.gattringer@uni-graz.at}\\
$ \quad \quad \quad \,  $ \email{ludovit.liptak@savba.sk}}

\abstract{ Recently a formulation of overlap fermions at finite 
density based on an analytic continuation of the sign function
was proposed. We study this proposal by analyzing the energy
and number densities for free fermions as a function
of the chemical potential and the temperature. Our results show that
overlap fermions with chemical potential give rise
to the correct continuum behavior.}

\FullConference{The XXV International Symposium on Lattice Field Theory\\
		 July 30 - August 4, 2007\\
		 Regensburg, Germany}

\begin{document}

\section{Introduction}

In recent years finite temperature and finite density physics has seen 
a lot of attention from the lattice QCD community. Although so far mainly
traditional fermion formulations such as staggered or Wilson fermions are
used, it is an important issue to get also exactly chiral fermions, in
particular the overlap operator \cite{overlap}, ready for QCD thermodynamics.  

In \cite{blwe} Bloch and Wettig have suggested a 
formulation of overlap fermions which includes a chemical potential.
The proposal is based on introducing the chemical potential as 
the 4-th component of the gauge field \cite{haka} and an analytic continuation
of the sign function into the complex plane, which requires new techniques for
the evaluation of the overlap operator \cite{bletal}. 

In a recent paper \cite{gali} we have studied thermodynamical properties of the
new proposal by analyzing the energy density for the free case as a function 
of the chemical potential and the temperature. In this contribution we augment
this analysis by computing also the number density and comparing it to the
known result for a free fermion gas. Our study indicates that the $\mu$- and 
$T$-dependence for both quantities, energy and number density, approach 
the continuum behavior correctly.

\section{Setup of the calculation}

The overlap Dirac operator $D(\mu)$ for
fermions with a chemical potential $\mu$ is given as \cite{blwe}
\begin{eqnarray}
D(\mu) &\; = \;& \frac{1}{a} [ 1 - \gamma_5 \, \mbox{sign}\, H(\mu) ] \; ,
\nonumber
\\
H(\mu) &\; = \;& \gamma_5 \, [ 1 - a D_W(\mu)] \; , 
\label{overlap}
\end{eqnarray}
where the sign function may be defined with the spectral 
theorem. $D_W(\mu)$ denotes the Wilson Dirac operator with chemical potential,
\begin{eqnarray}
D_W\!(\mu)_{x,y} &\; = \;& 
\mathds{1} \Big[\frac{3}{a} + \frac{1}{a_4} \Big] \, 
 \delta_{x,y} 
\\  
&\; - \; & 
\sum_{j = 1}^{3} \Big[ \,
\frac{\mathds{1}\! - \!\gamma_j}{2a} \, 
U_j(x) \, \delta_{x+\hat{j},\,y} \, + \,  
\frac{\mathds{1} \!+ \!\gamma_j}{2a} \,
U_j(x\!-\!\hat{j})^\dagger\, \delta_{x-\hat{j},\,y} \, \Big]   
\nonumber \\
& \; - \; & 
\frac{\mathds{1} \! - \!\gamma_4}{2a_4} \, 
U_4(x) \, e^{\mu a_4} \, \delta_{x+\hat{4},\,y} \, - \, 
\frac{\mathds{1} \!+\! \gamma_4}{2a_4} \,
U_4(x\!-\!\hat{4})^\dagger \, e^{-\mu a_4} \, \delta_{x-\hat{4},\,y} \; . 
\nonumber  
\end{eqnarray}
For later use we distinguish between the lattice spacing $a$ 
for the spatial directions (= 1,2,3 directions) and the temporal 
lattice constant $a_4$.
We use periodic boundary conditions in the spatial directions
and anti-periodic boundary conditions for time.
The chemical potential $\mu$ is introduced in the standard 
exponential form \cite{haka}.

For $\mu = 0$ the Wilson Dirac operator is $\gamma_5$-hermitian,
i.e.,\ $\gamma_5 D_W(0) \gamma_5 = D_W(0)^\dagger$, implying that $H(0)$
is a hermitian matrix. When the chemical 
potential $\mu$ is turned on, $\gamma_5$-hermiticity no longer holds, and 
$H(\mu)$ is a non-hermitian, general matrix. This fact implies: First,
the eigenvalues of $H(\mu)$ are no longer real and the 
sign function for a complex number has to be defined in the spectral 
representation of sign$\,H(\mu)$. Second, the spectral representation has
to be formulated using left- and right-eigenvectors. 
For the sign function of a complex number we use the analytic continuation 
proposed in ~\cite{blwe} and define the sign function through the sign of 
the real part
\begin{equation}
\mbox{sign}\, (x + i y) \; = \; \mbox{sign} \, (x) \; .
\end{equation}

For analyzing thermodynamic properties we study the 
following observables:
\begin{enumerate}
\item The energy density
\begin{eqnarray}
&& \epsilon(\mu) \; = \; \frac{1}{V} \langle {\cal H} \rangle \; 
= \; - \frac{1}{V} \frac{\partial}{\partial \beta} 
\ln \mbox{Tr} \Big[ e^{-\beta ( {\cal H} - \mu {\cal N})}\Big]_{\beta\mu = c}
\; = \; - \; \frac{1}{V} \frac{\partial \ln Z}{\partial \beta}
\bigg|_{\beta\mu = c} \; .
\label{epsidef}
\end{eqnarray}
\item The number density
\begin{eqnarray}
&& n_p (\mu) \; = \; \frac{1}{V} \langle {\cal N} \rangle \; = \; 
\frac{T}{V} \frac{\partial}{\partial \mu} 
\ln \mbox{Tr} \Big[ e^{-\beta ( {\cal H} - \mu {\cal N})}\Big]
\; = \;  \frac{T}{V} \frac{\partial \ln Z}{\partial \mu} \; .
\label{numidef}
\end{eqnarray}
\end{enumerate}
Here ${\cal H}$ is the Hamiltonian of the system, ${\cal N}$ 
denotes the number 
operator, and $\beta = 1/T$ is the inverse temperature (we set the 
Boltzmann constant equal to $1$). The derivatives 
in the definition of  $ \epsilon(\mu) $ are taken such that $\beta\mu = 
c =$ const. 

The continuum result for the subtracted thermodynamical quantities of
free massless fermions can be calculated using the definition 
(\ref{epsidef}) and (\ref{numidef}). The final expressions together 
with the logarithm of the partition function can be found in textbooks 
(see, e.g., ~\cite{kapusta}).

In the path integral formalism the inverse temperature $\beta$ is given by
the extent in 4-direction, which on the lattice has the form 
$\beta = N_4 a_4$. Thus the derivative $\partial/\partial \beta$ 
in (\ref{epsidef}) is expressed as
$N_4^{-1} \partial/\partial a_4$. The partition function $Z$ is given 
by the fermion determinant $\det D$ which we write as
the product over all eigenvalues $\lambda_n$ of $D(\mu)$. 
After some algebra one obtains 
\begin{eqnarray}
\epsilon(\mu) & \; =  \; & - \, \frac{1}{V N_4}
\frac{\partial \ln \det D}{\partial a_4}\bigg|_{a_4\mu = c}\!\!\!\! \!\!
= \; - \, \frac{1}{V N_4} \, \sum_n \, \frac{1}{\lambda_n} \; 
\frac{\partial \lambda_n}{\partial a_4}\bigg|_{a_4\mu = c}\;\;\; ,
\label{specsum}
\\
n_p (\mu) & \; = \;& \frac{1}{V N_4 a_4}
\frac{\partial \ln \det D}{\partial \mu} \; = \; \frac{1}{V N_4 a_4} \,
\sum_n \, \frac{1}{\lambda_n} \, 
\frac{\partial \lambda_n}{\partial \mu}\; .
\label{specsum1}
\end{eqnarray}

\section{Evaluation of the eigenvalues}

For the evaluation of (\ref{specsum}) and (\ref{specsum1}) we need the
eigenvalues $\lambda$ of $D(\mu)$ in closed form. This is done in three steps: 
First, using Fourier transformation,
we bring the Dirac operator for free fermions to $4\times4$ 
block-diagonal form, 
\begin{equation}
\widehat{H} \; = \; \gamma_5 \, h_5 \; + \; i \, \gamma_5 
\sum_\nu \, \gamma_\nu \, h_\nu 
\; ,
\end{equation}
with
\begin{eqnarray}
h_5 &\; = \;& 1 - \sum_{j=1}^3 [ 1 - \cos(a p_j)] - 
\frac{a}{a_4} [ 1 - \cos(a_4 (p_4 - i\mu))] \; ,
\nonumber \\
h_j &\; = \;& - \sin(a p_j) \quad \mbox{for} \quad j = 1,2,3 \; ,
\nonumber \\
h_4 &\; = \;& - \frac{a}{a_4} \sin(a_4 (p_4 - i\mu)) \; .
\label{hdefs}
\end{eqnarray}
The spatial momenta are given by $p_j =  2\pi k_j/aN$, where $N$ is the number 
of lattice points in the spatial directions and $k_j = 0,1\, ... \, N-1$.  
The momenta in time-direction are $p_4 =  \pi( 2k_4 + 1)/a_4 N_4$, 
$k_4 = 0,1 \, ...\, N_4-1$.

In the second step the spectral representation for the sign function of 
$\widehat{H}$ is constructed. The left- and right-eigenvectors of 
$\widehat{H}$, $l_j$ and $r_j$ are given by
\begin{eqnarray}
l_1 &\; = \;& l_1^{(0)} [ \widehat{H} + s \mathds{1} ] \;\; , \;\;   
l_2 \; = \; l_2^{(0)} [ \widehat{H} + s \mathds{1} ] \;\; , \;\;
l_3 \; = \; l_3^{(0)} [ \widehat{H} - s \mathds{1} ] \;\; , \;\;   
l_4 \; = \; l_4^{(0)} [ \widehat{H} - s \mathds{1} ] \; , 
\nonumber \\
r_1 &\; = \;& [ \widehat{H} + s \mathds{1} ] r_1^{(0)} \;\; , \;\;   
r_2 \; = \; [ \widehat{H} + s \mathds{1} ] r_2^{(0)} \;\; , \;\;   
r_3 \; = \; [ \widehat{H} - s \mathds{1} ] r_3^{(0)} \;\; , \;\;   
r_4 \; = \; [ \widehat{H} - s \mathds{1} ] r_4^{(0)} \; .
\end{eqnarray}
The constant spinors $l_j^{(0)}, r_j^{(0)}$ are 
($T$ denotes transposition)
\begin{eqnarray}
l_1^{(0)} & = & r_1^{(0)\, T} \, = \, c \, (1,0,0,0) \;\; , \;\;
l_2^{(0)} \, = \, r_2^{(0)\, T} \, = \, c \, (0,1,0,0) \, , 
\nonumber \\
l_3^{(0)} &\! = & r_3^{(0)\, T} \, = \, c \,(0,0,1,0) \;\; , \;\;
l_4^{(0)} \, = \, r_4^{(0)\, T} \, = \, c \, (0,0,0,1) \, .
\end{eqnarray} 
where $c = (2 s(s+h_5))^{-1/2}$ and $s = \sqrt{h^2 + h_5^2}$. 
The eigenvectors obey $l_i r_j = \delta_{ij}$.
The corresponding eigenvalues $\alpha_j$ of $\widehat{H}$ are
\begin{equation}
\alpha_1 = \alpha_2 \, = \, + \, s \; , \; 
\alpha_3 = \alpha_4 \, = \, - \, s \; .
\end{equation}
Using these eigenvalues and eigenvectors we can construct 
sign $\widehat{H}$ using the spectral 
\begin{equation}
\mbox{sign} \, \widehat{H} \; = \; 
\sum_{j=1}^4 \mbox{sign}\,(\lambda_j) \, r_j \, l_j \; = \;
\frac{\mbox{sign}(s)}{s} \, \widehat{H} \; .
\end{equation}

Inserting this result in (\ref{overlap}), the eigenvalues of the 
overlap operator at a given momentum are obtained as
\begin{equation}
\lambda_\pm \; = \; \frac{1}{a}\left[ 1 - 
\frac{ \mbox{sign}\,(\sqrt{h^2 + h_5^2}\,)\, h_5 \pm i 
\sqrt{h^2}}{\sqrt{h^2 + h_5^2}} \right] \; ,
\label{evals}
\end{equation}
where each of the two eigenvalues is twofold degenerate. 
In the spectral sums (\ref{specsum}), (\ref{specsum1}) 
the label $n$ runs over all
momenta and the eigenvalues at fixed momentum as given in (\ref{evals}).
The necessary derivative with respect to $a_4$ and 
$ \mu $ is straightforward to
compute in closed form, and the spectral sums 
(\ref{specsum}), (\ref{specsum1}) can then
be summed numerically. The argument of the sign function cannot become
purely imaginary
on a finite lattice, and no $\delta$-like terms occur. 
We remark, that after taking the derivative with respect to $a_4$, 
we set $a = a_4 = 1$, i.e., all the
results are presented in lattice units.

\section{Results}

We begin the discussion of our results with lattices which have equal length 
in all four directions, corresponding to zero temperature in the thermodynamic 
limit. We focus our attention on the subtracted energy density 
$\epsilon(\mu) - \epsilon(0)$  and the number density 
$n_p ( \mu ) - n_p ( 0 )$. The results for various lattice sizes are 
shown in Fig.~1.  According to \cite{kapusta}, we expect the 
data (symbols in Fig.~1) to approach the continuum form

\begin{equation}
\label{zerotempquan}
\epsilon ( \mu ) - \epsilon ( 0 ) = \frac{ \mu^4 }{4 \pi^2}  \quad , 
\quad n_p ( \mu ) - n_p ( 0 ) = \frac{\mu^3}{3 \pi^2} \; ,
\end{equation}
represented by dashed lines in Fig.~1.

We see clearly that both quantities show almost linear dependence on  
$\mu^4$ and $\mu^3$, respectively. The figure also demonstrates that 
with enlarging the lattice the finite size effects become smaller and 
the lattice results approach the continuum limit. The deviation from 
the linear dependence is due to the finite  extension of the lattice. 
Furthermore, we can expect that we can see finite temperature effects too.

\begin{figure}[t]
\label{figure2}
\begin{center}
\includegraphics[height=2.05in,clip]{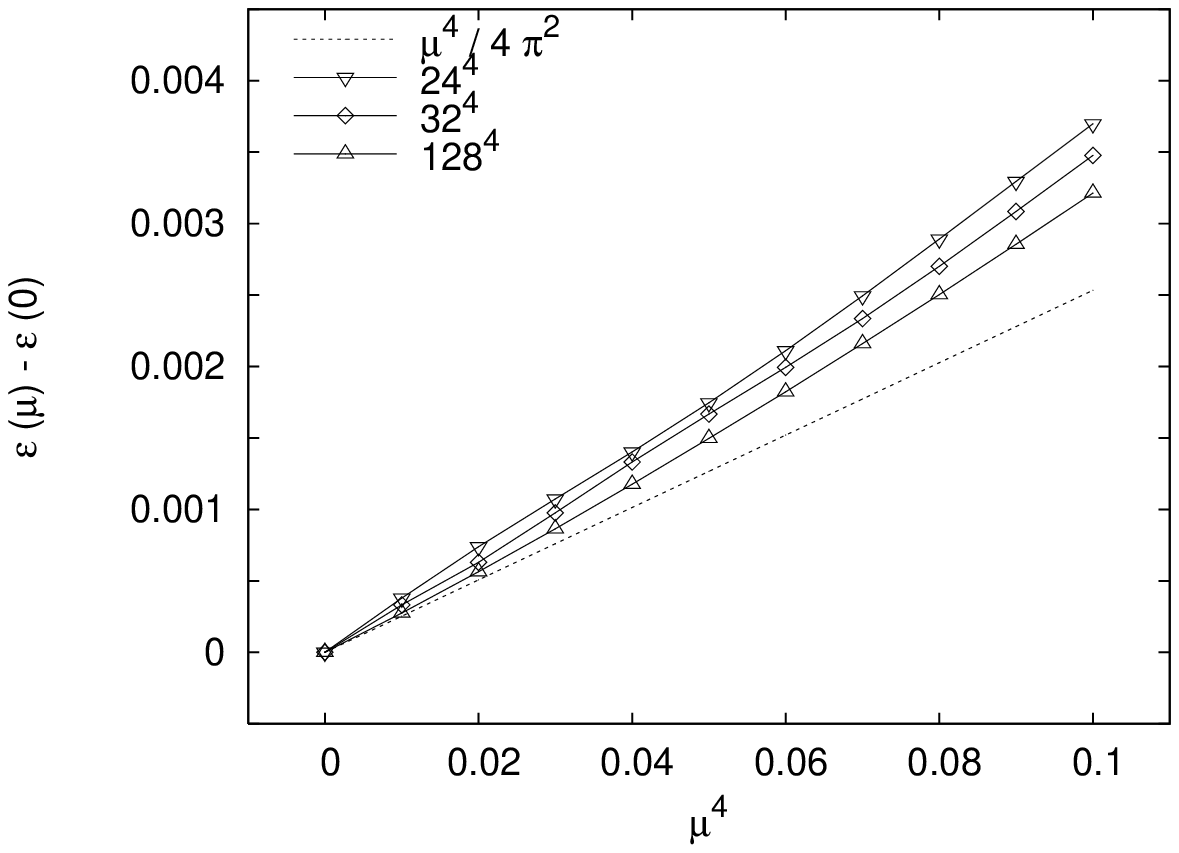}
\includegraphics[height=2.05in,clip]{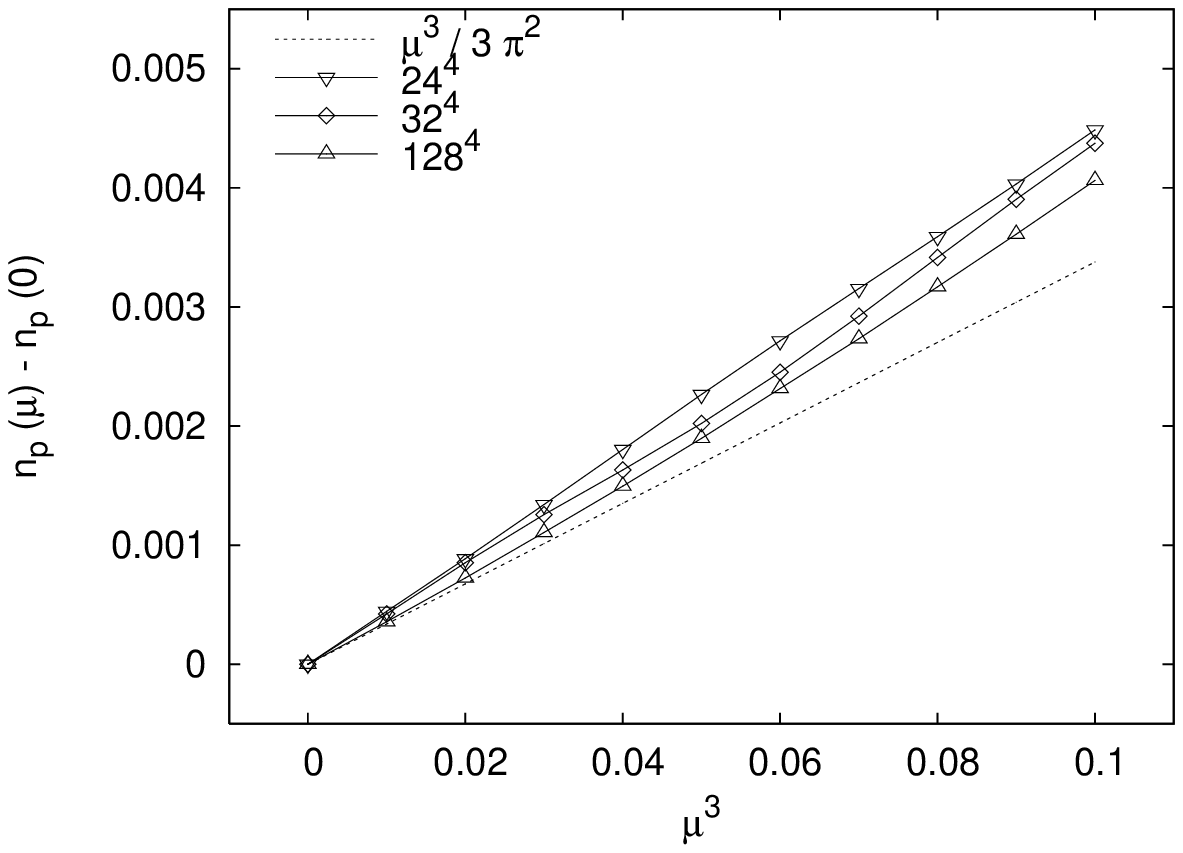}
\end{center}
\caption{The energy density $\epsilon ( \mu ) - \epsilon ( 0 )$  
and the number density  $ n_p ( \mu ) - n_p ( 0 ) $ as a function 
of $ \mu^4 $ and $ \mu^3 $, respectively. The symbols (connected to 
guide the eye) are for various lattice sizes, the dashed line is the 
continuum result. All quantities are given in lattice units.}
\end{figure}

To study the effects of temperature, we compare our data with the continuum
results \cite{kapusta} for the subtracted energy and the number density, now
including the leading temperature corrections,
\begin{equation}
\label{fintempquan}
\epsilon ( \mu , T ) - \epsilon ( 0 , T ) \; = \; 
\frac{ \mu^4 }{4 \pi^2} + \frac{T^2 \mu^2 }{2}  \quad , \quad 
n_p ( \mu , T ) - n_p ( 0 , T) \; = \; \frac{\mu^3}{3 \pi^2} + 
\frac{\mu T^2 }{3} \; .
\end{equation}

Fig.~2 shows the lattice results for finite temperature, using data from 
$128^3 \times N_4$ lattices with temperature $T = N_4^{-1}$ (in lattice units).
The approximate linearity of the zero-temperature case receives corrections as
the temperature is increased. The temperature-dependent terms have a 
square- and cubic-root behavior when we plot them as a function of 
$ \mu^4 $ and $ \mu^3 $, respectively.

\begin{figure}[t]
\label{figure3}
\begin{center}
\includegraphics[width=2.95in,clip]{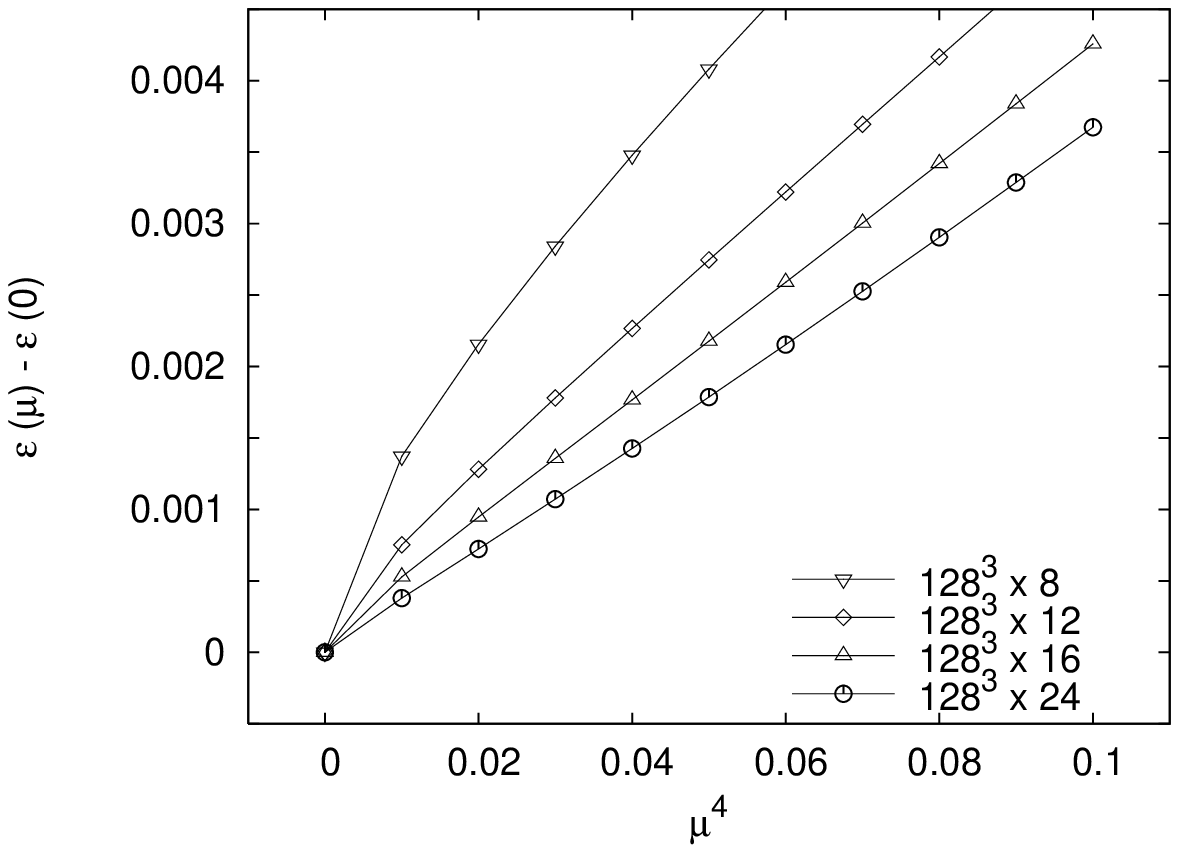}
\includegraphics[width=2.95in,clip]{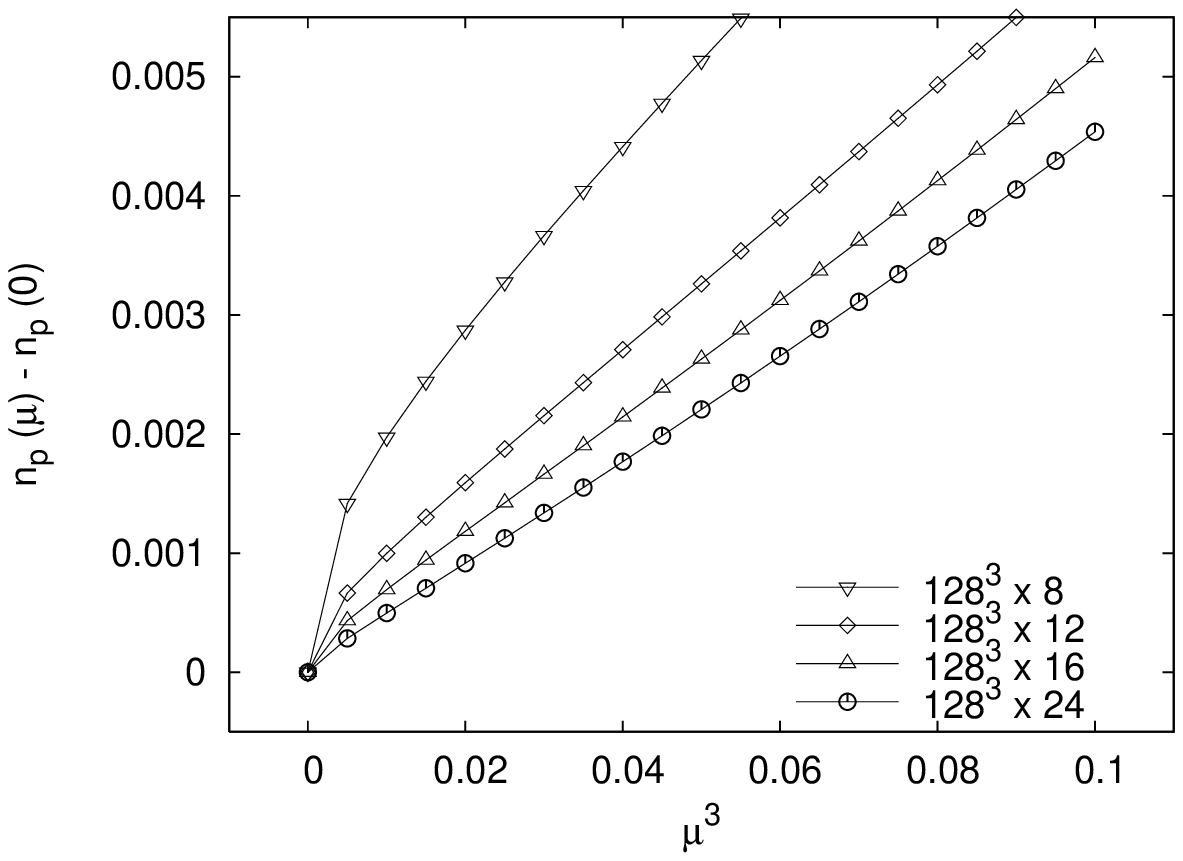}
\end{center}
\caption{The energy density $\epsilon ( \mu,T ) - \epsilon ( 0,T )$ 
and the number density $ n_p ( \mu,T ) - n_p ( 0,T ) $ as a function 
of $ \mu^4 $ and $ \mu^3 $, respectively, for finite temperature lattices.
All quantities are in lattice units.}
\end{figure}

In order to study the temperature effect quantitatively, we compare the finite 
temperature results to the continuum form (\ref{fintempquan}).  We have to 
take into account also finite size effects which we build in assuming 
contributions from higher terms of the expansion in $ \mu $. In the case 
of the energy density, we use the ansatz
\begin{equation}
\epsilon(\mu,T) - \epsilon (0,T) \; = \; 
c_2 \mu^2 \, + \, c_4 \mu^4 \, + \, c_6 \mu^6 \, + \, c_8 \mu^8 \, + \,
{\cal O}(\mu^{10})\; .
\label{fitfunction}
\end{equation}
For the number operator we use an equivalent form with odd powers of $ \mu $. 
The results of the fit for the data used in Fig.\ 2 and for the 
largest lattice of Fig.\ 1 are given in Table 1.

\renewcommand{\arraystretch}{1.1}
\begin{table}[b]
\begin{center}
\begin{tabular}{c|ccccc}
\hline
\hline
$N_4$  &  $N_4^{-2} / 2$ & $c_2$ & $c_4$ &  $D( c_2 )$ & $D( c_4 )$  \\
\hline
8    &  0.007812 & 0.010125 & 0.03519 & 29 \% & 39 \%  \\   
12   &  0.003472 & 0.004125 & 0.03178 & 19 \% & 25 \%  \\
16   &  0.001953 & 0.002192 & 0.02803 & 12 \% & 11 \%  \\
24   &  0.000868 & 0.000947 & 0.02587 &  9 \% &  2 \%   \\
128  &  0.000030 & 0.000032 & 0.02543 &  7 \% &  0.5 \% \\
\hline
\hline
\end{tabular}
\end{center}
\caption{Results for fits according to (4.3). 
The spatial volume is always $128^3$. The temporal extension $N_4$ is 
given in the first column. In the second column we list the 
corresponding value of $ N_4^{-2} / 2 $  which is what one expects 
for the fit coefficient $ c_2 $ in the third column.  
The coefficient $c_4$ is expected to approach the constant value 
$1 / 4 \pi^2 = 0.02533 $. $D(c_2)$ and $D(c_4)$  
are the relative deviations of $c_2$ and $c_4$ from the expected 
continuum values.}
\end{table}

As the table shows, the coefficients approach the continuum values
when the temporal extension is increased. In the case of the energy density, 
the discrepancy for $ N_4 = 24 $ is approximately $ 2 \% $ and $ 9 \% $ 
and for $ N_{128} $ only $ 0.5 \% $ and $ 7 \% $. For $N_4$ smaller than 24, 
the discrepancy for both coefficients is larger, which is due to 
finite size effects. Nevertheless both, the $\mu^4$- and $\mu^2 T^2$-terms,  
are well reproduced. The coefficients $c_6$ and $ c_8 $ have magnitudes of 
${\cal O}(10^{-2})$, 
so their contribution according to (\ref{fitfunction}) is irrelevant
for $ 0.1 < \mu$ (in lattice units). For the number density we find 
qualitatively the same behavior with only slightly different numerical values.

Finally, we analyze the $\mu$-dependence for large values of the chemical
potential, approaching the cutoff. In this case we compare our data with 
data obtained for the Wilson fermions and the continuum case. In Fig.~3 
we plot the ratio $\left(  \epsilon (\mu) - \epsilon (\mu) \right) / \mu^4$. 
The figure shows that in a certain interval, from zero up to approximately 
$0.7$ in lattice units, both lattice formulations give practically identical 
results and start to differ only beyond this already
relatively large value of $\mu$. 
The second observation is that up to $ \mu \approx 0.6$ the lattice results 
differ very little from the continuum form and discretization errors 
grow large only for $\mu \gtrsim 0.6$. Also in a comparison with the data for 
fermions obtained by blocking \cite {biwi} one again finds good
agreement. 

\begin{figure}[t]
\begin{center}
\includegraphics[width=3.5in]{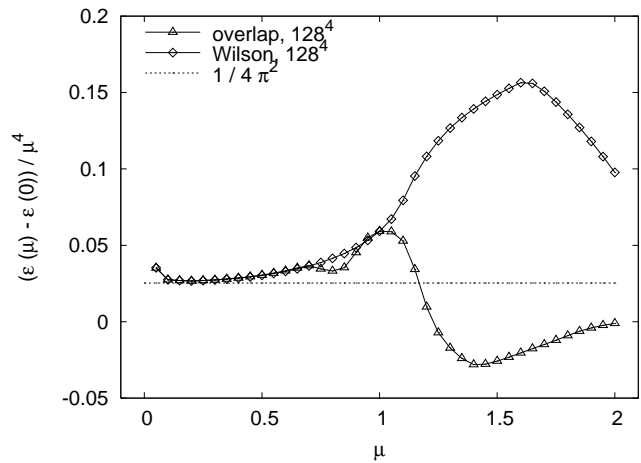}
\end{center}
\caption{The ratio of $\left( \epsilon (\mu) - \epsilon(\mu) \right) / \mu^4$ 
as a function of $\mu$. The figure compares overlap fermions with Wilson 
results and the continuum.}
\end{figure}

\section{Summary}

We have analyzed thermodynamical quantities for the overlap operator at 
finite chemical potential obtained by analytic continuation of 
the sign function 
into the complex plane as suggested in \cite{blwe}. The analysis was done 
for finite and zero temperature lattices. Fits of the data show that the 
expected continuum behavior is reliably approached and no trace of 
unphysical $\mu^2/a^2$ terms was found. We conclude that the suggested 
introduction of chemical potential \cite{blwe} provides both, 
chiral symmetry and the correct description of fermions at finite density.

\end{document}